\documentclass[prb,showpacs,showkeys,preprintnumbers,superscriptaddress,a4paper,twocolumn]{revtex4-1}
\usepackage[T1]{fontenc}
\usepackage[utf8]{inputenc}
\usepackage{graphicx}
\usepackage{amsmath,amssymb}
\usepackage{amstext}
\usepackage{lipsum}
\usepackage{mathtools}
\usepackage{afterpage}
\usepackage{epsfig}
\usepackage{epstopdf} 
\usepackage{filecontents}
\usepackage[colorlinks,citecolor=blue,filecolor=blue,linkcolor=blue,urlcolor=blue]{hyperref}
\usepackage{braket}
\usepackage{dsfont}
\usepackage{color}
\usepackage[dvipsnames]{xcolor}
\usepackage{mathrsfs}
\usepackage{setspace}
\usepackage{caption}
\usepackage{enumerate}
\usepackage{nicefrac}
\usepackage{bm}
\usepackage{subfig}
\usepackage{ragged2e}
\DeclareCaptionJustification{justified}{\justifying}
\makeatletter
\newcommand{\DC}[0]{\ext@arrow 0100\leftrightarrowfill@{}{\text{DC}}}
\makeatother
\newcommand{\myvec}[1]{\textbf{#1}}

\DeclareMathOperator{\sign}{\text{sign}}

\begin{document}
\title{Existence of nodal line semi-metal in a generalized three dimensional Haldane model }
\author{Sudarshan Saha}\email{sudarshan@iopb.res.in}\affiliation{Institute of Physics, Bhubaneswar- 751005, Odhisa, India}
\affiliation{Homi Bhabha National Institute, Mumbai - 400 094, Maharashtra, India}
\author{Saptarshi Mandal}\email{saptarshi@iopb.res.in}\affiliation{Institute of Physics, Bhubaneswar- 751005, Odhisa, India}
\affiliation{Homi Bhabha National Institute, Mumbai - 400 094, Maharashtra, India}
	
	\begin{abstract}

We construct and study a time reversal broken tight binding model on diamond lattice with complex next-nearest-neighbour hopping which
can be thought of as a generalisation of two dimensional Haldane model in three dimension. The model also breaks inversion symmetry owing to sub-lattice dependent chemical potential. We calculate the spectrum of the model and find the existence of six pairs of anisotropic gapless points with linear dependence on momentum. The  coordinates of the gapless points are ($2 \pi, \pi \pm k_0,0),~ (2 \pi, \pi \pm k_0,2 \pi)$ and their possible permutations . The condition for gapless spectrum is very similar to the two dimensional case. Each gapless points are having well defined chirality and in the gapless phase specific set of planes have non-zero Chern number. The gapped phase is a trivial bulk insulator  which has vanishing Chern number as well as Hopf index. The model  belongs to the symmetry class AIII according to the ten-fold way of classification.  Surprisingly the gapless phase does  contain a gapped surface state where as the gapped state has a gapless  surface states as found in (1,1,1) direction. 

\end{abstract}

\keywords{3-d topological insulator; hopf invariant, Pontryjagen invariant, hopf insulator}
	
\date{\today}
	
\pacs{Insert PACS number here}
\maketitle
	
\section{Introduction}
	
   Understanding and classification of materials in the  group of metal and insulator has been a long time interest 
to the scientific endeavour.  Starting from the simple prediction of Bloch theorem, we have come across a long 
way examining the role of electron-electron interaction, disorder, impurity, pressure, spin-orbit coupling, temperature 
etc. In this context, a very intimate and interesting connection of topology with the conductivity tensor has been 
established in early 80's for the integer quantum hall effect(IQHE)\cite{thouless-1982}. It has been shown that conductivity 
tensor is related to Chern number for the problem of electron gas in the presence of magnetic field which breaks the time 
reversal symmetry explicitly. Few year later,  Haldane showed that external magnetic field is not necessary 
condition to have quantized  Hall conductivity though time reversal breaking is a necessary ingredient \cite{haldane-1988}.
A fundamental jump in this direction happened when Kene and Mele showed that time reversal symmetry breaking is not a necessary
ingredient to have a topological insulating phase~\cite{kane-2005-2nd}. The model includes a next-nearest neighbour 
spin-orbit coupled hopping on honeycomb lattice. It has been shown that a non-zero quantum spin hall current can be
linked to the same Chern number. However very soon it has been realized that  this simple classification on the basis of 
Chern number is insufficient for the case of time reversal invariant system. For $\mathcal{T}$ invariant system, a Z2 classification
has been proposed  as the simple Chern number must reduce to zero for $S_z$ non-conserving terms are included in the form of Rashbha
coupling~\cite{kane-2005-1st}. \\

  Very soon the proposed $Z_2$ classification was extended to 3 Dimensional time reversal invariant quantum spin hall insulator(QSHI) \cite{fu-kane-2007,moore-2007} in terms
of four $Z_2$ invariants resulting 16 classes topological insulators. A connection of $Z_2$ invariants with the Chern invariants was also shown \cite{rahul-2009-1,rahul-2009-2} for the time reversal symmetric(TRS) system. On the other hand in 3 Dimension for time reversal and inversion symmetry broken system~\cite{halperin-1987,kohmoto-1992,ezzine-2004,guo-2015}
has yielded the Weyl systems with non-trivial connection to surface states to bulk gapless mode~\cite{armitage-2018,murakami-2007,wan-2011,hosur-2012} . These Weyl fermionic systems are natural generalizations  relativistic Dirac fermions in 3 Dimensions and topological invariants are the Chirality
which is determined by calculating the Chern number around the gapless Dirac node over a closed surface. However all these models have a four component
structure in momentum space. For two component Hamiltonian in 3 dimension it is realized that a different kind of Topological insulator named
Hopf Insulator exist which is characterized by Hopf invariants which is particularly useful when the other topological invariant named so far
is not directly usable in particular when the Chern number becomes zero in 3 Dimension \cite{moore-2008,deng-2013}. These motivates us to extend the Haldane model in three dimension in the Diamond lattice which contains next-nearest complex hopping as chosen
in Haldane model. The model involves a spinless fermion hopping in  diamond lattice and hamiltonian has only $2 \times 2$ structure for each momentum due to two sub-lattice structure of diamond lattice. Thus it fills an important chapter in the study of three dimensional topological system owing to its absence of both time reversal and inversion symmetry and an effective $2 \times 2 $ structure.\\

   The paper has been organized in the following manner. In Sec. \ref{sec:extensionOfHaldaneModel} we define the model and its  band structure and low energy expansion in momentum space. In Sec.  \ref{sec:symmetry} we discuss the symmetry of the model   according to the ten-fold way classification \cite{altland-1997,zimbauer-1996,andreas-2008}. In Sec. \ref{sec:surfaceState} numerical calculation of surface state is discussed. In Sec. \ref{chno} we discuss the topology of the model by determining the Chern number and Hopf invariant and discuss its connection with Weyl physics. Finally we conclude our study in  sec. \ref{sec:conclusion}.\\

\section{Extension of Haldane model to 3-dimension}
\label{sec:extensionOfHaldaneModel}
\begin{figure}
\begin{minipage}{0.1\textwidth}
\includegraphics[scale=0.40]{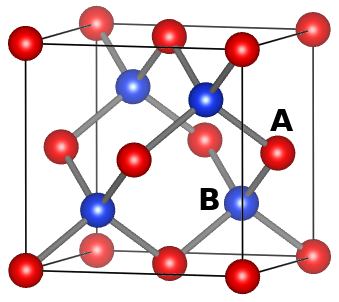}
\end{minipage}\hfill
\begin{minipage}{0.23\textwidth}
\includegraphics[scale=0.80]{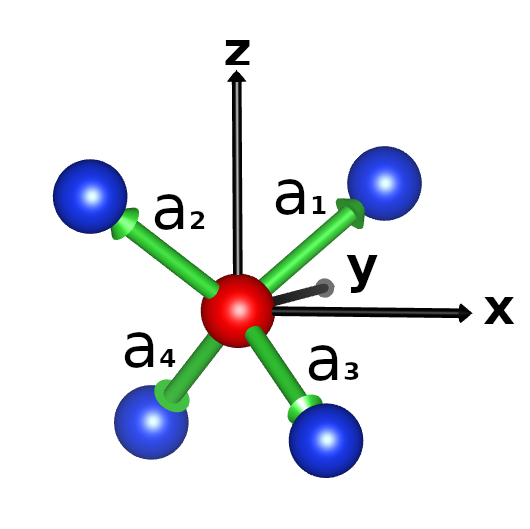}
\end{minipage}	
\caption{In the left panel we show a conventional cubic cell of the diamond lattice. Two inter-penetrating FCC sub-lattice are shown by red and blue dots which corresponds to `A' and `B' respectively. In the right panel, we show the four
nearest neighbour vectors that has been used in the text.}
\label{fig:diamondLattice}
\end{figure}

We have chosen the diamond lattice for the extension of Haldane model in 3 dimension for its similarity with two dimensional honeycomb lattice in having  hexagonal plaquette as an elementary unit. Diamond lattice has two inter-penetrating FCC sub-lattice denoted by  `A' and `B' as shown in the  Fig. \ref{fig:diamondLattice}.   The Model Hamiltonian we are interested in this study is following,
\begin{eqnarray}
H &=& t_1\sum\limits_{\langle ij\rangle}b_i^\dagger a_j + t_2\sum\limits_{\langle\langle ij\rangle\rangle}e^{i\nu_{ij}\phi}a_i^{\dagger}a_j + M\sum\limits_ia_i^{\dagger}a_i\notag\\
&& + t_2\sum\limits_{\langle\langle ij\rangle\rangle}e^{-i\nu_{ij}\phi}b_i^{\dagger}b_j - M\sum\limits_ib_i^{\dagger}b_i + \text{h.c.}
\label{eqn:nnnHamiltonian}
\end{eqnarray}
	In the above  $t_1$ denotes nearest neighbour (NN) hopping and $t_2$ denotes next nearest neighbour (NNN) hopping between the same sub-lattice. We have taken into account a complex phase with $t_2$ in Haldane sense which breaks the time reversal symmetry explicitly. Here  $\nu_{ij}$ has following properties, $\nu_{ij} = -\nu_{ji}$, $\nu_{-i-j} = -\nu_{ij}$. Details of NN and NNN-vectors are given in the Appendix \ref{co-ord}. After Fourier transformation in pseudo-spin space $\psi_{k}=(a_k, b_k)$ Hamiltonian is given by
\begin{eqnarray}
H&=& \sum_k \psi^{*}_k h_k \psi_k
\end{eqnarray}
where $h_k$ is a $2\times2$ matrix and is given by,
\begin{eqnarray}
h_{\myvec{k}}& =& 2t_2 F(k)\mathds{1} + 4t_1 C(k)\sigma_x -  4 t_1 S(k)\sigma_y \nonumber \\
&&+  2t_2\sin(\phi)(G(k) + M )\sigma_z
\label{eqn:hk_NNN_sigmaMatrix}
\end{eqnarray}
The various components $h_{ij}(k)$ that appear in $h_k$ are given below with the convention $h_{11}(k)=h_{1k}, h_{22}(k)=h_{2k}, h_{12}(k)=h_{3,k}=h^*_{21}(k)$ ,
\begin{eqnarray}
&&h_{1k} = 2t_2 \left[ F(\myvec{k}) \cos\phi + G(\myvec{k}) \sin\phi \right] + M \\
\label{eqn:h_11}
&&h_{3k} = h^*(k)_{12} = 4t_1(C(k)-i S(k) \\
&& h_{2k} = 2t_2 ( F(\myvec{k}) \cos\phi - G(\myvec{k}) \sin\phi )\\
&& F(\myvec{k}) = \sum_{\beta = 1}^{6}\cos(\myvec{k}\cdot\myvec{b}_\beta), G(\myvec{k}) = \sum_{\beta = 1}^{6}\sin(\myvec{k}\cdot\myvec{b}_\beta) \\
&&C(k)=\cos^2(k_x/4)\cos^2(k_y/4)\cos^2(k_z/4) \\
&&S(k)=\sin^2(k_x/4)\sin^2(k_y/4)\sin^2(k_z/4)
\label{eqn:h_22}
\end{eqnarray}

The dispersion obtained after diagonalization is given below, 
\begin{equation}
\mathcal{E}(k)_\pm = \frac{1}{2}\left[ (h_{1k} + h_{2k}) \pm \sqrt{(h_{1k} - h_{2k})^2 + 4|h_{3k}|^2}\; \right]
\label{eqn:E}
\end{equation}
From this dispersion relation (Eq-$\!$ \ref{eqn:E}) gap closing condition are obtained  as follows,
\begin{equation}
( h_{1k} - h_{2k})^2 = 0 = h_{3k} 
\label{eqn:firstCondition}
\end{equation}
From Eq-$\:$\ref{eqn:h_11} and Eq-$\:$\ref{eqn:h_22} first condition translates into
\begin{equation}
M + 2t_2\sin\phi\sum\limits_{\beta = 1}^{6}\sin(\myvec{k}\cdot\myvec{b}_\beta) = 0
\label{eqn:secondCondition.2}
\end{equation}

\begin{figure}
\centering
\includegraphics[scale=0.25]{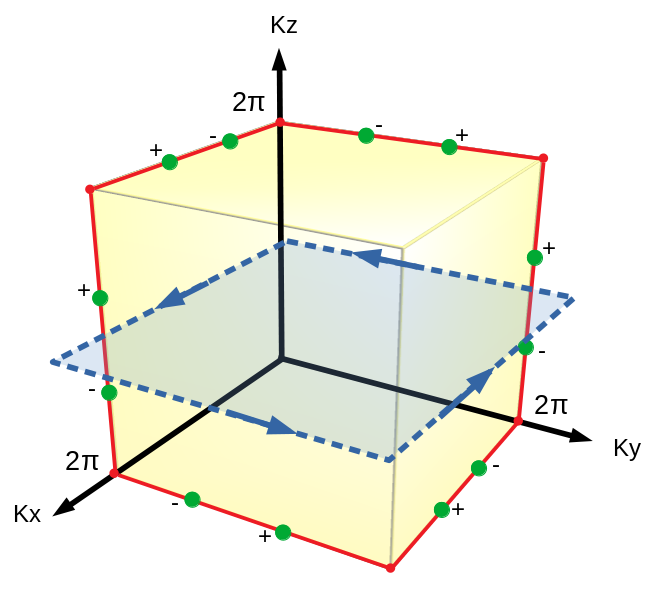}
\caption{Nodal line of nearest-neighbour tight-binding model on diamond lattice is the orange contour. For the Hamiltonian of the model, the green points are anisotropic  Weyl like points with well defined chirality.  The blue plane is one of those plane for which Chern number is calculated in Sec. \ref{chno}.}
\label{fig:weylPoints}
\end{figure}
\begin{figure}[!h]
\includegraphics[width=0.4\textwidth]{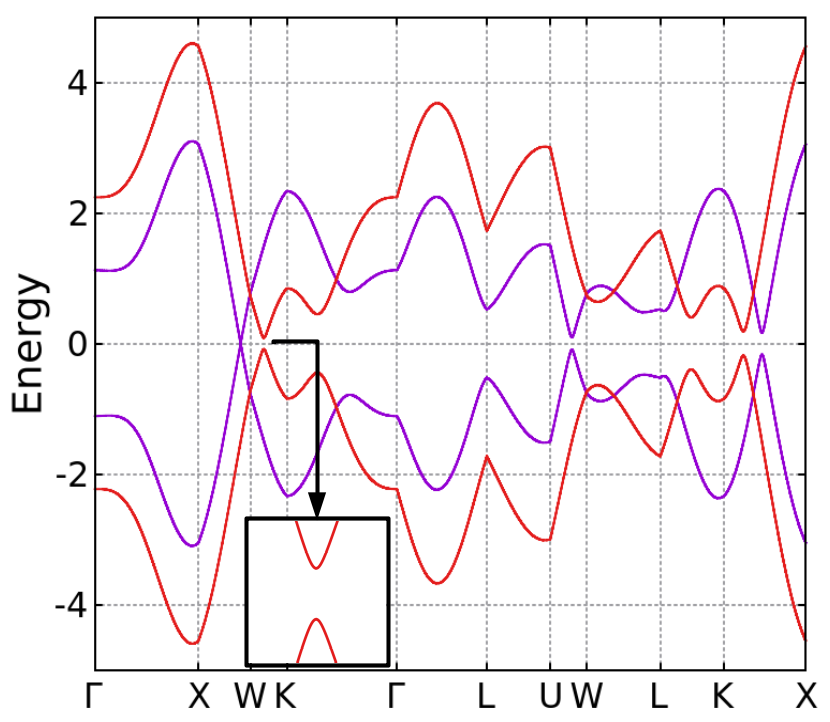}
\caption{In the above, the violet plot is the spectrum for the parameter $M=0.50, \Phi=1.57,$ and for the red plot $M=2.0, \Phi=1.57$. For
both the plot $t_1=1.0, t_2=0.5$. We notice that the gapless point occurs away from the high symmetry point.}
\label{fig:bandStructure1}
\end{figure}
\begin{figure}
\includegraphics[scale=0.15]{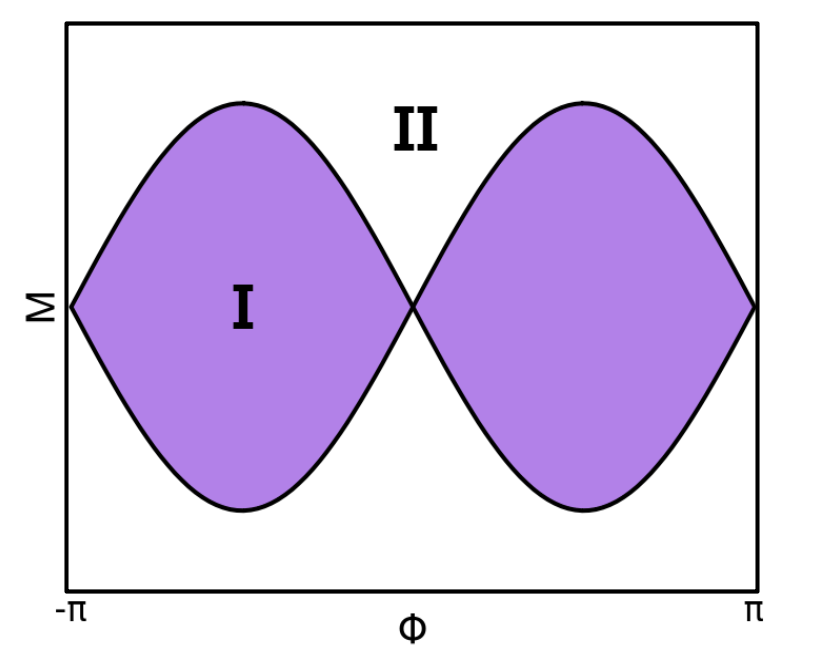}
\caption{The above plots shows the region of gapless phase and gapped phase as denoted by I and II respectively for  $t_1=1, t_2=0.5$. }
\label{fig:gaplessregion}
\end{figure}

Now the condition $h_{3k}=0$ enforces the gapless points to  reside on the nodal line  having equation $k_{\alpha/\beta}=0, 2\pi,  \alpha=x,y,z$. We have six such nodal lines as  presented in Fig. \ref{fig:weylPoints}. For simplicity, we consider the gap closing point
on the nodal line having  $k_x = 2\pi$ and $k_y = 0$  yielding  $\sum_{\beta = 1}^{6}\sin(\myvec{k}\cdot\myvec{b}_\beta) = 2\sin(k_z/2)$ and from Eq-$\:$\ref{eqn:secondCondition.2} we  obtain the following condition for gaplessness,
\begin{equation}
M \pm  4t_2\sin(k_z/2)\sin\phi = 0
\label{eqn:secondCondition.3}
\end{equation}
Similar condition for gaplessness can be obtained by replacing $k_z$ to $k_{x/y}$. The above condition for gaplessness can be compared with two dimensional case where one finds 
$M=\pm 3\sqrt{3} t_2 \sin \phi$.  From the Eq. \ref{eqn:secondCondition.3}, it is clear that  for a given $M$ and $\phi$ we can find $k_z$ from Eq-$\!$ \ref{eqn:secondCondition.3} at which gap closes. In Fig. \ref{fig:gaplessregion}, we present the regions where the gap closes or not in $M-\phi$ plane. The region-I denotes the  gapless phase and region-II denotes the gapped phase. We notice that gapless points occurs at the isolated points on the nodal line only as shown in Fig. \ref{fig:weylPoints}.  Now it is customary to obtain the low energy effective Hamiltonian around these gapless points. A straightforward calculation starting from  Eq-$\!$ \ref{eqn:hk_NNN_sigmaMatrix} around the nodal line $(k_x = 0, k_y = 2\pi)$ and for any $k_z$ satisfying Eq. \ref{eqn:secondCondition.3}, we obtain the effective Hamiltonian as,

\begin{eqnarray}
&&\mathcal{H}_{(0,2\pi,k_{z0})}= h_{0} \mathcal{I} + \sum \sigma_i \nabla_{\mu}h_{i,k} \delta k^{\mu}, ~~i=x,y,z 
\label{weylpoints}
\end{eqnarray}
The various component appearing in the above expressions are as follows,

\begin{eqnarray}
\label{weylcomponent1}
&&h_0 = -4t_2\cos\phi,~~ h_{x,k} = -t_1\cos(k_{z0}/4) q_{y}=h_{y,k},  \\
&& h_{z,k} =  2t_2\sin\phi \left[ \cos(k_{z,0}/2)\left\{ q_x - q_y - q_z \right\} - q_x \right]
\label{weylcomponent2}
\end{eqnarray}
In the above $k_{z,0}$ is the values of $k_z$ for which gapless condition is satisfied according to Eq. \ref{eqn:secondCondition.3}.
From Eq. \ref{weylpoints}, Eq. \ref{weylcomponent1} and Eq. \ref{weylcomponent2}, it is clear that this is an asymmetric Weyl like points. The x-component does depend on the y-component and the z-component is linear combination of more than one component. The label $(0,2\pi,k_{z0})$  denotes the points at which the above low energy expansion has been obtained. We notice that there are $12$ such points and similar expansion can be obtained around each Wyel point. \\
\indent
 Generically each such gapless points are characterized by their Chirality  $(C)$ and it is given by $C = \sign(\myvec{v}_x\cdot\myvec{h}_y\times\myvec{v}_z)$\cite{armitage-2018} where $\myvec{h}_i = \bm{\nabla_k}h_i, \; i = x, y, z$ and $\myvec{h}$ is given in Eq.~\ref{weylpoints}, Eq.~\ref{weylcomponent1} and Eq.~\ref{weylcomponent2}. For example, Chirality of gapless point at $(2\pi, 0, \pi + \delta k)$ is $\sign(\phi\:\delta k)$. The Chirality of each gapless points which has also been depicted in Fig. \ref{fig:weylPoints}.

\section{Symmetry}
\label{sec:symmetry}
	Following Altland and Zirnbauer\cite{altland-1997}, Hamiltonian of non-interacting fermionic systems can be classified on the basis of three basic discrete symmetries such as Time-reversal symmetry $\mathcal{T}$ (anti-unitary and commutes with the Hamiltonian), Particle-hole symmetry $\mathcal{P}$ (anti-unitary and anti-commutes with the Hamiltonian) and chirality symmetry $\mathcal{C}$ (unitary and anti-commutes with the Hamiltonian). In our study we implement the $\mathcal{P}$ symmetry by the transformation  $a_i \leftrightarrow b_i^\dagger$ and the inversion symmetry $\mathcal{I}$ $a_i \leftrightarrow b_{-i}$, to the  Hamiltonian in Eq-$\!$ \ref{eqn:nnnHamiltonian}. For clarity we denote $H_{ab}$ as the nearest neighbour hopping part of the Eq. \ref{eqn:nnnHamiltonian}. Similarly $H_{aa}$ and $H_{bb}$ denote the next-nearest neighbour hopping interaction and $H_{M}$ denotes the sub-lattice dependent chemical potential term.  These various parts of the Hamiltonian undergo the following transformation under $\mathcal{P}$ followed by $\mathcal{I}$.
\begin{eqnarray}
&& H_{ab} \xrightarrow{\mathcal{P}, \mathcal{I}} - H_{ab} \\
&& H_{aa} \xrightarrow{\mathcal{P}, \mathcal{I}} - H_{bb} \\
&& H_{bb} \xrightarrow{\mathcal{P}, \mathcal{I}} - H_{aa} \\
&& H_{M} \xrightarrow{\mathcal{P}, \mathcal{I}} - H_{M} 
\end{eqnarray}
The above mapping indeed tells that Hamiltonian in Eq. \ref{eqn:nnnHamiltonian} gets an overall negative sign. This symmetry is unitary and anti-commutes with the Hamiltonian and according to ten-fold way of classification it falls in the AIII Chiral classes and this class is characterized by the $\mathbb{Z}$, infinite cyclic group which  must be characterized by some integers.

\section{Surface state}
\label{sec:surfaceState}
\begin{figure}
\includegraphics[scale=0.35]{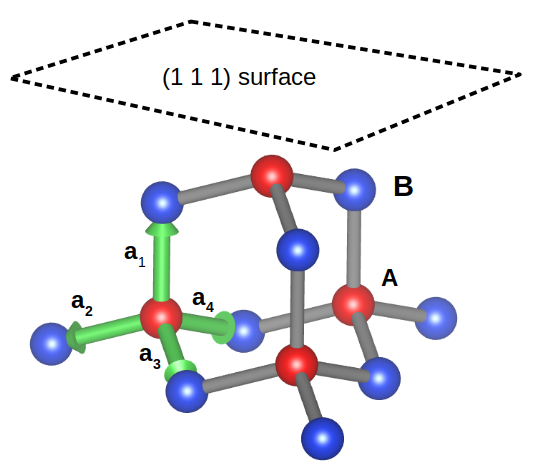}
\caption{Schematic of diamond lattice for slab geometry with (111) as the boundary surface perpendicular to $\myvec{a}_1$ NN-vector. A (Red) and B (Blue) atoms denote the two sub-lattices.}
\label{fig:surfaceStateGeometry}
\end{figure}

\begin{figure}[!htb]
  \includegraphics[width=.45\linewidth]{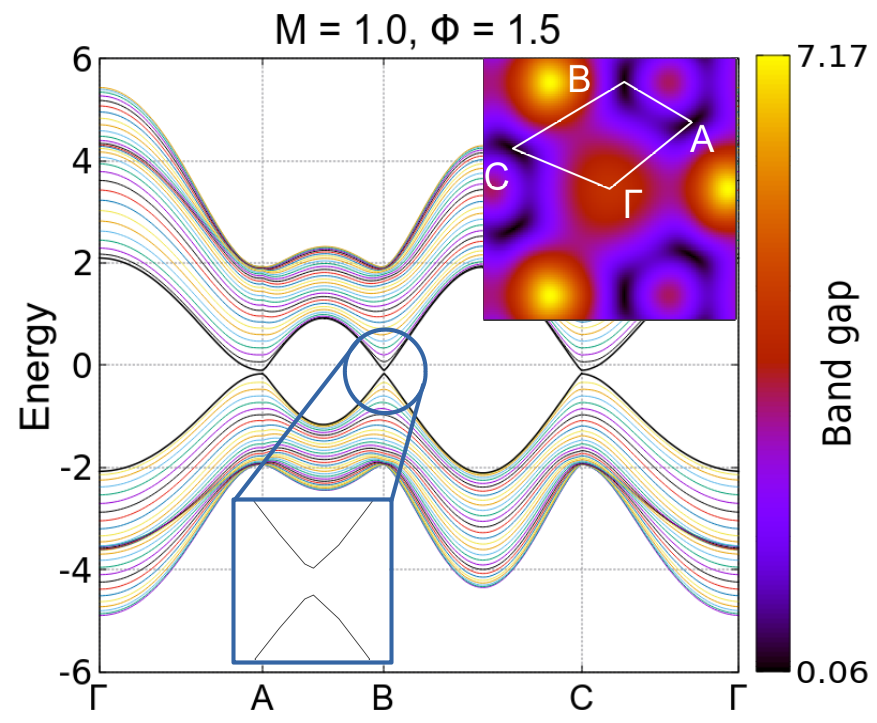}
  \includegraphics[width=.45\linewidth]{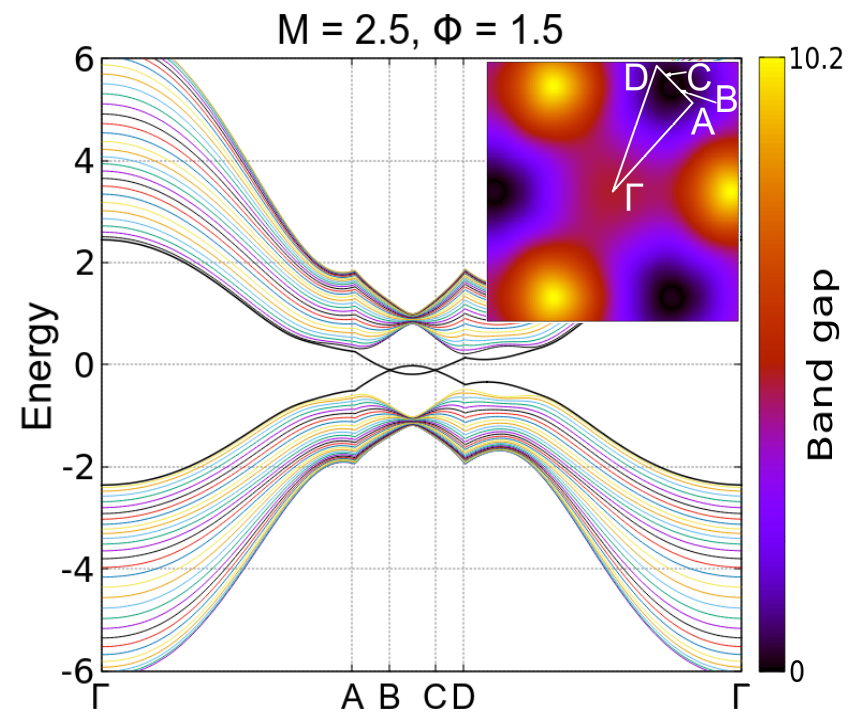}
\caption{In above figures we have shown the spectrum for an open surface geometry and taken 30 layers. The lowest positive eigenvalue corresponds to
surface state in (1,1,1) direction. In the left panel we have shown the spectrum for gapless region with $M=1.0, \Phi=1.5$. The surface state is gapped here. In the right panel we have shown the spectrum for gapped region and in this phase the surface state  is gapless. The minimum of the energy for the surface state does not occur on the contour joining the high symmetry points. The exact location of the points (A,B,C) and (ABCD) that appears in the left and right panel of this figure has been given in the left panel of Fig. \ref{fig:surfacegapped}.}
\label{fig:bandStructureNNN}
\end{figure}

Having discussed the spectrum of the model in details, the existence of the asymmetric Weyl like points along the nodal line and necessary symmetry classification, we discuss the nature of surface states. To this end, we calculate the surface state numerically of our model 
for a slab geometry with the boundary surface in (111) direction, as shown in the Fig.$\!$ \ref{fig:surfaceStateGeometry}.  We have found that the
gapless phase where the bulk is gapless do contain a gapped surface states as shown in left panel of Fig.  \ref{fig:bandStructureNNN}. 
In the right panel of Fig. \ref{fig:bandStructureNNN}, we show the band structure for the same open geometry in gapped phase and find 
the existence of a pair of gapless surface states that crosses each other twice. In the Fig. \ref{fig:bandStructureNNN}, we have plotted the band structure in two dimensional- Brillioune zone along a certain direction which deviates from the high symmetry points. 



\begin{figure}[!h]
\includegraphics[width=.30\linewidth]{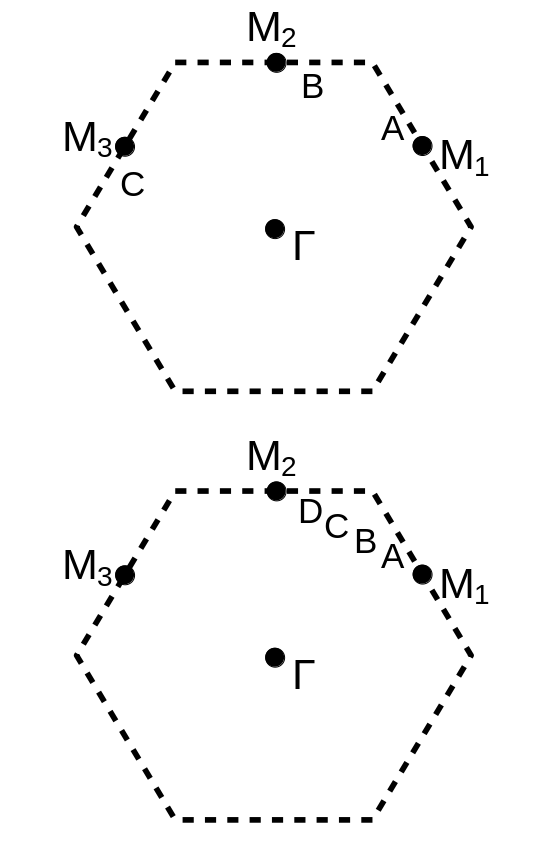}
\includegraphics[scale=0.25]{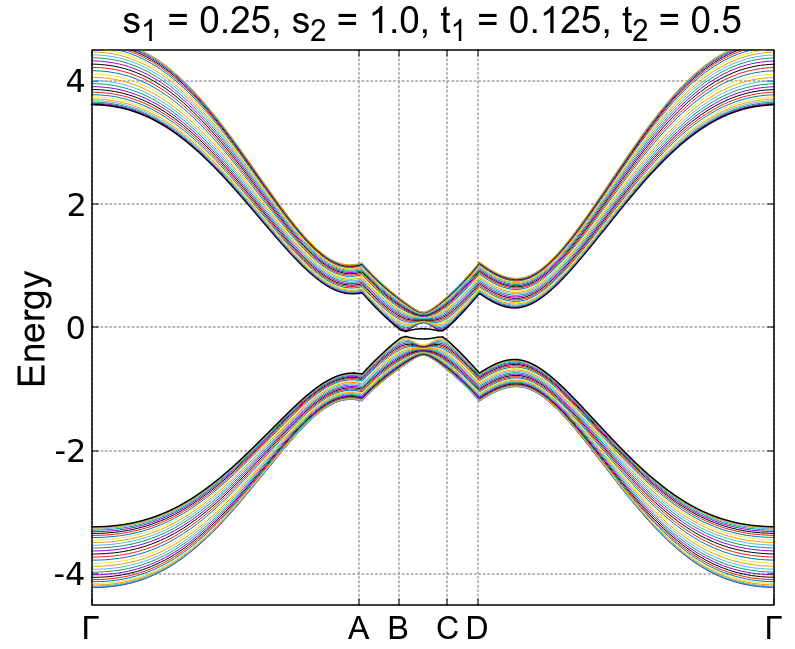}
\caption{The band structure for finite slab geometry with (1,1,1) surface open has been shown for the parameter values mentioned above the figure and explained
in the text. We see that for these parameter values, the surface state is gapped.}
\label{fig:surfacegapped}
\end{figure}

\begin{figure}[!htb]
  \includegraphics[width=.45\linewidth]{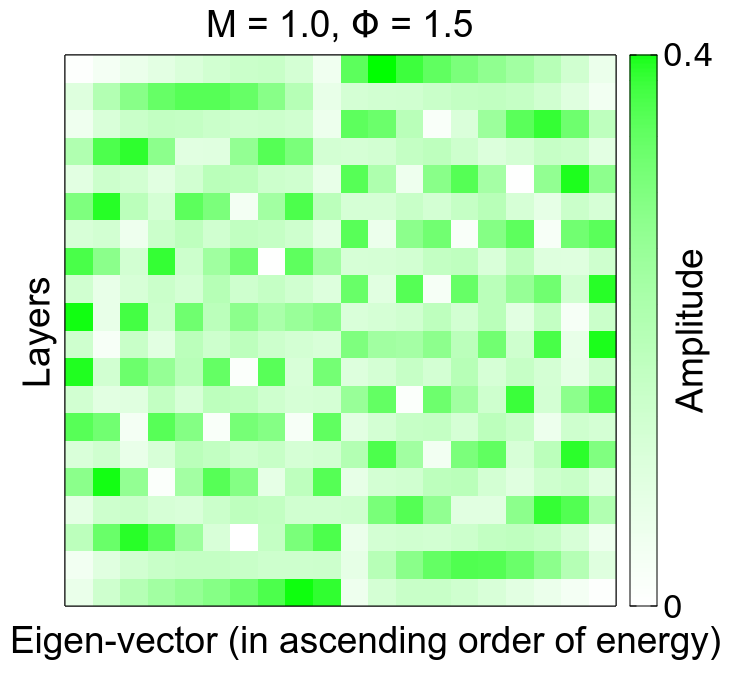}
  \includegraphics[width=.45\linewidth]{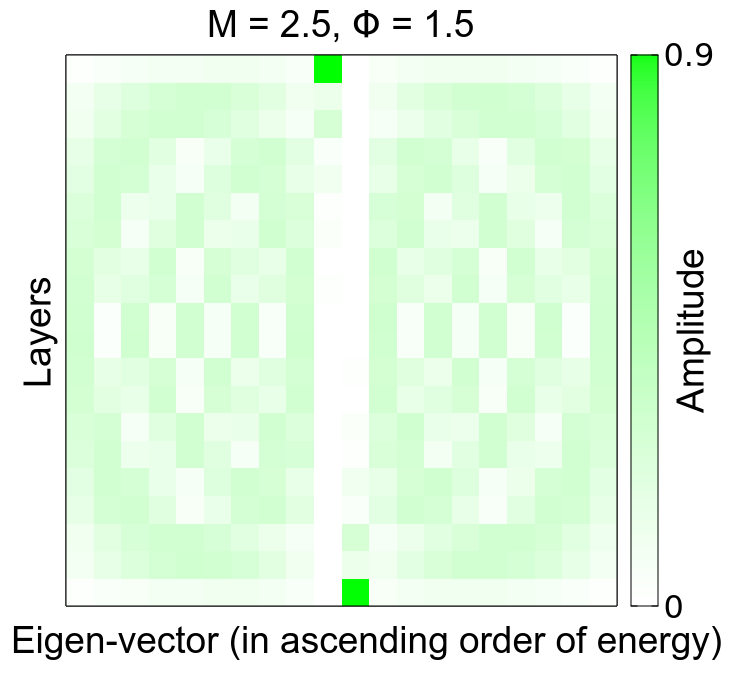}
\caption{In above figures we have shown how the amplitude of eigenfunctions corresponding to a given momentum `$k$' as explained in the text.  The amplitude corresponds to Fig.~\ref{fig:bandStructureNNN} and we notice that in the gapped region (left panel), the surface state  decays rapidly in comparison to the surface state in gapless region(right panel.)}
\label{fig:surfaceState}
\end{figure}

This has been done to include the gapless points along the contour which does not lie on the contour connecting the high symmetry points in the present model. 
The  gapless surface state that exist in the gapped region can be made gapped by changing the  parameter values. One easy way to do this is to change the ratio of interlayer and 
intra layer hopping amplitude. Let $s_1$ and $t_1$ denotes the interlayer and intra layer nearest-neighbour hopping. Similarly $s_2$ and $t_2$ denotes the interlayer and 
intra layer hopping amplitude. For certain combinations of $s_i$ and $t_i$, the gap can be opened for the surface state in gapped region. In Fig. \ref{fig:surfacegapped}
we have shown a gapped surface state in gapped region which happens for $s_1=0.25, s_2=1.0, t_1=0.125, t_2=0.5$. Such transition from gapless surface state to gapped surface state can experimentally detected by surface conductance.
In Fig.~\ref{fig:surfaceState} we show absolute value of component of eigenvectors as a heat-map showing how the  surface-state appears and  decays into bulk starting from one of the open surface. In drawing this figure we have used a projection of one Weyl like points on the two dimensional brillouin zone in the gapless phase. For the gapped phase `$k$' is the point `B' in  the right panel of Fig.~\ref{fig:bandStructureNNN} which is one of the gap closing point for the surface state. In conclusion we have found always a \textit{gapped surface state} in Region I and \textit{gap-less surface state} in Region II which can be made gapped by changing the parameter  values.
	
\section{Chern number and Hopf invariant}
\label{chno}
Unlike two dimensional topological insulators, the topological classification of three Dimensional systems are more involved. For the time reversal invariant system generally $Z_2$ index is used to classify the topology where as for time reversal broken system Chern number can still be used. In the absence of finite Chern no, it has been proposed that Hopf index could be finite and can be used to characterize the system.
Though we are dealing with a three dimensional system, substantial amount of information can be extracted from the topological property of two dimensional sub-system which constitute the whole system ~\cite{rahul-2009-1,rahul-2009-2,rahul-2007}. Though the connection of Chern number for three dimensional system has been discussed in the context of time reversal invariant system we deduce those for our case as first hand probe to the nature of  topology of our system. We note that for the original Haldane model time reversal is broken though there were no external magnetic field was present and it is the Chern number which decided the topological nature of the phases. This also motivates us to derive the Chern number for our system. Since all three directions $k_x$, $k_y$ and $k_z$ are equivalent, we analyse only for $k_z$ direction and same conclusion  holds for all other directions. Let's consider a set of planes perpendicular to the $k_z$ axis which is intercepted by two nodal lines $(k_x = 0, \; k_y = 2\pi)$ and $(k_x = 2\pi, \; k_y = 0)$. In Fig. \ref{fig:weylPoints}, the square blue shaded region denotes one such plane. To calculate Chern number for such planes we consider three cases i) $k_z=0, 2 \pi, \rm{any~arbitrary~value}$. We are going to consider the third case first. Depending on the choice of parameters $(M, \phi)$ if $H_z(\myvec{k})$ changes sign in the plane we can not have single gauge choice throughout and the Berry vector potential in these two region is connected by the relation $A_I=A_{II} + \nabla \psi$ where $\psi$ is obtained as,
\begin{eqnarray}
e^{i\psi(\myvec{k})} &=& \dfrac{|h_x(k) + i h_y(k)|}{h_x(k) + i h_y(k)} \nonumber \\
  &=& \dfrac{|\beta \cos\theta - i\alpha \sin\theta|}{\beta \cos\theta - i\alpha \sin\theta} \nonumber \\
  &=& e^{i\nicefrac{\pi}{2} - i\eta(\theta)} 
\end{eqnarray}
In the above, we have used   $h_x(k)$, $h_y(k)$, following Eq.~\ref{weylcomponent1} and Eq.~\ref{weylcomponent2} with $h_z(k)= 2 t_2 \sin \phi(G(k)+M)$.  Using short notation as $\alpha = \cos(k_z/4)$, $\beta = \sin(k_z/4)$, and  $q_x = q\cos\theta$, $q_y = q\sin\theta$ and $\tan\eta = \tfrac{\alpha}{\beta}\tan\theta$ . From the above equation,  one can show that $ \psi(\myvec{k}) = \dfrac{\pi}{2} - \eta(\theta)$. Now, since $0<k_z< 2\pi$, Chern number $\mathcal{C}$ for this case is,
\begin{eqnarray}
\mathcal{C}=\tfrac{1}{2\pi}\oint d\myvec{k}\cdot\bm{\nabla}_k\psi(\myvec{k}) = -1
\end{eqnarray}
 
Similarly for the first case, $k_z = 0$, this plane has two consecutive sides along which gap closes. On this plane we have, $C_k = - q_y,~ S_k=0$. So the phase difference $\psi$ is given by, $e^{i\psi(\myvec{k})} = \dfrac{|-t_1q_y|}{-t_1q_y}$. Hence the Chern no can be found as ,
\begin{equation}
\mathcal{C}=\psi(2\pi + \epsilon) - \psi(\epsilon) = 0 \qquad (\epsilon-\text{small})
\end{equation}

The above results can be reconciled from the condition in Eq. \ref{eqn:secondCondition.3} that for $k_z=0$, the gapless condition reduces to $M=0$ and
$\phi$ independent. The above discussion can be summarized in the following way. For the gapless phase as denoted by region-I in Fig. \ref{fig:surfaceState}, for every plane parallel to $k_x-k_y$ plane and having a definite $k_z$ in between the two Weyl-like points, there is a non-zero Chern number for a given $\phi, M$. Thus we have found continuous set of parallel plane for which Chern number is non-zero. The above arguments is true for other planes parallel to $k_x-K_z$ and $k_z-K_y$ plane.  In the gapped phase there is no plane having non-zero Chern no. This construction is very similar  found earlier \cite{rahul-2009-1,rahul-2009-2}. \\
\noindent	

From the analysis of our model Hamiltonian, we have found that the gapless surface state exist for the region where bulk is gapped. On the contrary, the region where bulk is gapless we have gapped surface state.  We have also shown in the preceding discussion that there are six planes for which non-zero chern number exist. The gapless phase contains anisotropic Weyl-like points and it is known that generally in such case, there exist a gapless surface states for open  systems~\cite{kundu,sumathi-rev}. However our results is in contrast to the existing literature.  To explore more on the topology theory we note that it has been established from the homotopy theory
that for Hamiltonian having $2\times2$ structure  in 3 dimension and in the absence of finite Chern number, the Hopf invariant can take finite integer and this signifies a non-trivial topology in 3 dimension \cite{moore-2008,deng-2013}. It may be noted that for 2 dimension a $2\times 2 $ Hamiltonian in momentum space is a map from $T^2 \rightarrow S^2$. However in 3 dimension a $2 \times 2$ Hamiltonian is a map from $T^3 \rightarrow S^3 \rightarrow S^2$. The Hopf invariant is associated with the possible non-triviality of the map $S^3 \rightarrow S^2$. If the Chern numbers $C_\mu = 0$ in all three directions, the Hopf index takes all integers values $\mathbb{Z}$ and has a simple integral expression
\begin{equation}
\chi(\hat{\myvec{u}}) = -\int_{\text{BZ}}\myvec{F}\cdot\myvec{A}\;d\myvec{k}
\label{hopfinvariant}
\end{equation}
	where $\myvec{A}$ is the Berry connection which satisfies $\myvec{F} = \bm{\nabla}\times\myvec{A}$ and $\myvec{u}$ denotes the map from $T^3 \rightarrow S^2$ which in our case is the wave function. The Hopf index $\chi(\hat{\myvec{u}})$ is gauge invariant although its expression depends on $\myvec{A}$. The Hopf invariant is similar to the $\mathbb{Z}_2$ invariant or Chern parity in topological insulators in that its standard integral expression uses the gauge-dependent quantity $\myvec{A}$ even though the final result is gauge-invariant when the Chern numbers are zero. In our gapped regime where a non-trivial topological surface state exist and the Chern number is zero, we have  evaluated numerically the l.h.s of Eq. \ref{hopfinvariant} and found to be zero. Thus  unlike the other topological systems \cite{moore-2008,deng-2013} which have Chern number zero but finite Hopf index, the gapped phase we find here seems to be trivial bulk insulator.  \\
\section{Conclusion}
\label{sec:conclusion}
	In essence we have studied an extension of two dimension Haldane model \cite{haldane-1988} in diamond lattice providing a 3 dimensional generalisation. Both time reversal and inversion symmetry has been broken with complex coupling constant and sub-lattice depended mass term respectively, but keeping chiral symmetry intact. Topological classification has been performed on the basis of dispersion relation, Chern number of two-dimensional sub-systems and existence of surface state in the parameter space $(M,\phi)$ has been done. In some region of $(M, \phi)$ space the system is gapless only at certain number of $\myvec{k}$ points and we find existence of gapless surface state at $\myvec{k}$ where system is gapped. In the gapless phase there exist 6 pairs asymmetric Weyl-like points with well defined chirality. In the gapless state the surface state is always gapped but in the gapped phase the surface state is gapless or gapped depending on the parameter values.  We have calculated Hopf index to characterize this gapped phase but that turnes out to be zero.  The insulating phase thus seems to be a trivial bulk insulator.\\

\section*{Acknowledgement} The authors acknowledg many fruitful discussion with Arijit Saha and Kush Saha.
	
\section{Appendix-A}
\label{appendix}
\subsection{Details co-ordinate system used}
\label{co-ord}
Here we first enlist the nearest-neighbour vectors used.
\begin{eqnarray}
\vec{a}_1 &=& \frac{1}{4} (1,1,1) \nonumber \\
\vec{a}_2 &=& \frac{1}{4} (-1,1,-1) \nonumber \\
\vec{a}_3 &=& \frac{1}{4} (-1,-1,1) \nonumber \\
\vec{a}_1 &=& \frac{1}{4} (1,-1,-1) 
\end{eqnarray}

The details of next-nearest-neighbour vectors are given below.
\begin{eqnarray}
\vec{b}_1 &=& \frac{1}{2} (1,0,1) \nonumber \\
\vec{b}_2 &=& \frac{1}{2} (1,1,0) \nonumber \\
\vec{b}_3 &=& \frac{1}{2} (0,1,1) \nonumber \\
\vec{b}_4 &=& \frac{1}{2} (0,1,-1) \nonumber \\
\vec{b}_5 &=& \frac{1}{2} (1,-1,0) \nonumber \\
\vec{b}_1 &=& \frac{1}{2} (-1,0,1) 
\end{eqnarray}
\subsection{surface state}
In Sec. \ref{sec:surfaceState}, we have presented the numerical results for the surface state where the system has been kept open in (1,1,1)
direction as shown in Fig. \ref{fig:surfaceStateGeometry}. Here we give detail derivation of the equations for such geometry and derive the equations which needs to be solved to obtain the surface states analytically~\cite{pereg-2005,you-2008,huang-2017} .  Following reference \cite{wilf}, we can easily derive the  Harper equation for this
slab geometry by defining a Fourier transformation in a two dimension plane equivalent to  a honeycomb lattice. The coupled equation that one obtains are given below where $a(k,n)$ and $b(k,n)$ represent that annihilation operator at `A' and `B' sub-lattice  and $k$ and $n$ refers to momentum and layer index respectively. 

\begin{widetext}
\begin{equation}
\begin{split}
a(k,n+1)t_2e^{i\phi}[1 + e^{ik_1} +{} & e^{ik_2}] + a(k,n)[2t_2\{  \cos(k_1 + \phi) + \cos(k_2 + \phi) + \cos(k_3 + \phi) \} + M]\\ & + a(k,n-1)t_2e^{-i\phi}[1 + e^{-ik_1} + e^{-ik_2}] + b(k,n)t_1[1 + e^{ik_1} + e^{-ik_3}] + b(k,n-1)t_1 = 0
\label{eqn:firstRecurssionRelation}
\end{split}
\end{equation}
\begin{equation}
\begin{split}
a(k,n+1)t_1 +{} & a(k,n)t_1[1 + e^{-ik_1} + e^{ik_3}] + b(k,n+1)t_2e^{-i\phi}[1 + e^{ik_1} + e^{ik_2}]\\ & + b(k,n)[2t_2\{ \cos(k_1 - \phi) + \cos(k_2 - \phi) + \cos(k_3 - \phi) \} - M] + b(k,n-1)t_2e^{i\phi}[1 + e^{-ik_1} + e^{-ik_2}] = 0
\label{eqn:secndRecurssionRelation}
\end{split}
\end{equation}
\end{widetext}
	
where $k_1 = \bm{\kappa}\cdot(\myvec{a}_2 - \myvec{a}_3)$, $k_2 = \bm{\kappa}\cdot(\myvec{a}_2 - \myvec{a}_4)$ and $k_3 = \bm{\kappa}\cdot(\myvec{a}_3 - \myvec{a}_4)$ and $\myvec{a}_2$, $\myvec{a}_3$, $\myvec{a}_4$ are shown in the figure \ref{fig:surfaceStateGeometry}. The above set of coupled linear homogeneous recursion relation can be solved by \textit{method of generating function}> For this purpose we define two functions $f(z)$ and $g(z)$ in the following way
\begin{equation*}
f(z) = \sum\limits_{n=0}^{\infty} a(n)z^n, \quad	g(z) = \sum\limits_{n=0}^{\infty} b(n)z^n
\end{equation*}
where $a(n)$, $b(n)$ are solution of the recursion relations \ref{eqn:firstRecurssionRelation}, \ref{eqn:secndRecurssionRelation} and explicite dependence on `$k$' has been
omitted for simplicity. The formal solution for the coefficient $a(n)$ and $b(n)$ are obtained as,
\begin{align}
a(n) &= \left.\dfrac{1}{n!}\dfrac{d^n}{dz^n}f(z)\right|_{z=0} = \dfrac{1}{2\pi i}\oint\dfrac{f(z)}{z^{n+1}}dz
\label{eqn:a_n}\\
b(n) &= \left.\dfrac{1}{n!}\dfrac{d^n}{dz^n}g(z)\right|_{z=0} = \dfrac{1}{2\pi i}\oint\dfrac{g(z)}{z^{n+1}}dz
\end{align}
and $f(z)$, $g(z)$ turns out to be
\begin{widetext}
\begin{align}
f(z) &= \frac{\bar{\alpha}\alpha z +(\bar{\beta}\alpha - \delta t_1 )z^2 + (\bar{\alpha}^*\alpha - t_1^2)z^3}{\alpha\bar{\alpha} + [\alpha\bar{\beta} + \beta\bar{\alpha} - \delta t_1]z + [\alpha\bar{\alpha}^* + \alpha^*\bar{\alpha} + \beta\bar{\beta} - \delta\delta^* - {t_1}^2]z^2 + [\beta\bar{\alpha}^* + \alpha^*\bar{\beta} - t_1\delta^*]z^3 + \alpha^*\bar{\alpha}^*z^4}
\label{eqn:fz}\\
g(z) &= \frac{-\alpha\delta^*z^2 + \alpha^*t_1z^3}{\alpha\bar{\alpha} + [\alpha\bar{\beta} + \beta\bar{\alpha} - \delta t_1]z + [\alpha\bar{\alpha}^* + \alpha^*\bar{\alpha} + \beta\bar{\beta} - \delta\delta^* - {t_1}^2]z^2 + [\beta\bar{\alpha}^* + \alpha^*\bar{\beta} - t_1\delta^*]z^3 + \alpha^*\bar{\alpha}^*z^4}
\label{eqn:gz}
\end{align}
\end{widetext}
where
\begin{eqnarray}
\alpha& =& \gamma^* = t_2e^{i\phi}[1 + e^{ik_1} + e^{ik_2}]\\
\beta &=& \sum\limits_{i=1}^{3} 2t_2 \cos(k_i + \phi) + M \\
\delta& =& t_1[1 + e^{-ik_1} + e^{-ik_2}]
\end{eqnarray}
	In the above, $\bar{\alpha}(M, \phi) = \alpha(-M, -\phi)$, $\bar{\beta}(M, \phi) = \beta(-M, -\phi)$ and $(\ast)$ denotes complex conjugation. The boundary condition we have used is $a_0 = 0$, $a_1 = 1$ and $b_0 = b_1 = 0$. But this does not affects later analysis since it appears only in the numerator. The above set of equation as given in Eq. \ref{eqn:fz}
and Eq. \ref{eqn:gz} can be solved using the method of residue. However it turns out that it is not very straightforward to obtain the solution and it will be presented elsewhere

\end{document}